# Évaluation et consolidation d'un réseau lexical via un outil pour retrouver le mot sur le bout de la langue

Alain Joubert (1), Mathieu Lafourcade (1), Didier Schwab (2), Michael Zock (3)

(1) LIRMM, Université Montpellier II   (2) LIG, Université Grenoble II   (3) LIF-CNRS, Marseille
{alain.joubert, mathieu.lafourcade}@lirmm.fr, didier.schwab@imag.fr, michael.zock@lif.univ-mrs.fr

**Résumé** Depuis septembre 2007, un réseau lexical de grande taille pour le Français est en cours de construction à l'aide de méthodes fondées sur des formes de consensus populaire obtenu via des jeux (projet JeuxDeMots). L'intervention d'experts humains est marginale en ce qu'elle représente moins de 0,5% des relations du réseau et se limite à des corrections, à des ajustements ainsi qu'à la validation des sens de termes. Pour évaluer la qualité de cette ressource construite par des participants de jeu (utilisateurs non experts) nous adoptons une démarche similaire à celle de sa construction, à savoir, la ressource doit être validée sur un vocabulaire de classe ouverte, par des non-experts, de façon stable (persistante dans le temps). Pour ce faire, nous proposons de vérifier si notre ressource est capable de servir de support à la résolution du problème nommé 'Mot sur le Bout de la Langue' (MBL). A l'instar de JeuxdeMots, l'outil développé peut être vu comme un jeu en ligne. Tout comme ce dernier, il permet d'acquérir de nouvelles relations, constituant ainsi un enrichissement de notre réseau lexical.

**Abstract** Since September 2007, a large scale lexical network for French is under construction through methods based on some kind of popular consensus by means of games (JeuxDeMots project). Human intervention can be considered as marginal. It is limited to corrections, adjustments and validation of the senses of terms, which amounts to less than 0,5 % of the relations in the network. To appreciate the quality of this resource built by non-expert users (players of the game), we use a similar approach to its construction. The resource must be validated by laymen, persistent in time, on open class vocabulary. We suggest to check whether our tool is able to solve the *Tip of the Tongue* (TOT) problem. Just like JeuxDeMots, our tool can be considered as an on-line game. Like the former, it allows the acquisition of new relations, enriching thus the (existing) network.

**Mots-clés**   Réseau lexical, JeuxDeMots, évaluation, outil de MBL, mot sur le bout de la langue

## Introduction

Grâce à un nombre important de participants à des jeux en ligne (notamment JeuxDeMots et PtiClic), nous avons obtenu un réseau lexical de grande taille pour la langue française (actuellement plus de 220000 termes[1], reliés par plus d'un million de relations sémantiques) représentant une connaissance générale commune. La communauté dispose donc d'une ressource lexicale dont nous souhaitons évaluer la qualité. Une évaluation manuelle pose au moins deux problèmes : d'une part, elle peut être biaisée par les compétences de l'évaluateur, et d'autre part, elle nécessite un temps prohibitif dès que l'on souhaite effectuer une évaluation quelque peu conséquente. Nous aurions pu envisager une évaluation automatique par comparaison avec une référence, mais à notre connaissance une telle référence n'existe pas, du moins

---
[1] Un terme peut être constitué de plusieurs mots (par exemple : *étoile de mer*)

pour la langue française, ayant une couverture et un nombre de types de relation suffisant. L'évaluation manuelle par échantillonnage ne nous semble pas satisfaisante car elle est nécessairement trop réduite, trop ponctuelle et d'une qualité difficile à apprécier. Nous avons donc décidé d'évaluer notre ressource, via un logiciel de détermination du « mot sur le bout de la langue » (MBL), évaluation qui pourrait être faite de façon permanente et avec un grand nombre d'évaluateurs en simple aveugle (ces derniers ne sachant pas qu'ils évaluent). Notre réseau lexical représentant des connaissances générales, un tel logiciel doit s'appliquer à un domaine ouvert, à savoir du vocabulaire tout venant, y compris des termes peu courants. Compte tenu du caractère sémantique du réseau lexical, l'outil de MBL opèrera exclusivement de manière quasi-sémantique, utilisant essentiellement des associations d'idées, des relations ontologiques ou celles de typicalité. Un mode d'accès par la phonétique, voire la notion de rébus, est donc exclu.

Nous commencerons cet article en présentant d'abord la problématique du MBL, pour rappeler ensuite brièvement le processus de constitution de notre réseau lexical, avant de présenter notre outil de MBL, dénommé AKI[2]. Enfin, nous commenterons les résultats obtenus grâce à AKI pour évaluer notre réseau lexical.

## 1   Problématique

**Difficulté de l'évaluation -** Nous sommes confrontés au problème d'évaluation de donnée lexicale, où aucun standard de référence n'est disponible et où une l'évaluation manuelle n'est pas envisageable. Dans un premier temps, on serait tenté de répondre aux questions de complétude et de précision (exactitude) :

- notre réseau lexical est-il « complet », à savoir comporte-t-il « tous » les termes et « toutes » les relations ?
- n'y a-t-il pas dans notre réseau des termes ou des relations erronés ?

Bien évidemment, la réponse stricte à ces deux questions est négative, ne serait-ce qu'en raison du caractère évolutif de la langue ; par exemple, le terme *révolution de jasmin* n'est apparu qu'en janvier 2011. Cependant, nous pouvons dégager une question plus réaliste :

- pour chaque terme de notre réseau lexical, l'ensemble des relations qu'il entretient avec d'autres termes suffit-il à le caractériser de façon unique ?

Dans l'affirmative, tout terme est susceptible d'être retrouvé via un ensemble réduit de termes indices. Ceci étant, nous avons créé un outil de "mot sur le bout de la langue" (MBL) pour réaliser cette évaluation.

## 2   Le problème du 'mot sur le bout de la langue' (MBL)

Le terme "manque de mot" désigne à la fois l'absence de terme dans le dictionnaire mental (Aitchison, 2003) d'un locuteur, ainsi que l'incapacité de pouvoir y accéder à temps. Nous nous intéressons ici uniquement à ce dernier cas. Le manque de mot, est une situation connue par tout producteur de langue notamment à l'oral (discours spontané). Cette défaillance sera qualifiée d'anomie, d'Alzheimer, ou de *mot sur le bout de la langue* (MBL), selon la durée et la fréquence du blocage et selon la nature d'information accessible (sémantique, phonologique) au moment crucial, la production écrite ou orale.

L'expression *« avoir le mot sur le bout de la langue »* (MBL) » ou, son analogue anglais, *« it's on the tip of my tongue »* (TOT), décrivent une forme de blocage très particulier. Un locuteur cherchant à exprimer une idée est conscient de connaître le terme, il sent sa production imminente (le plus gros du travail étant accompli), pourtant, il échoue tout près de la fin. La dernière partie est inaccessible. Tel un éternuement non consommé, la forme sonore reste bloquée, et la traduction du sens en forme linguistique n'aboutit pas. Ce qui caractérise le MBL et ce qui le distingue d'autres formes de manque de mot[3] c'est que le locuteur connaît

---

[2]   http://www.lirmm.fr/jeuxdemots/AKI.php
[3]   Comme indiqué, il y a d'autres cas de figure d'échec lexical. L'un, où le locuteur ignore tout simplement le mot recherché (cas fréquent en langue étrangère), et l'autre, où il connaît le terme, mais il n'arrive pas à l'évoquer à temps : c'est le blanc ou le vide total, situation typique pour des mots rares ou très techniques. D'ailleurs, beaucoup de gens utilisent le terme de MBL de manière générique, voulant lui faire endosser tout type de manque de mot. Ceci est impropre, car il peut y avoir différentes raisons pour

le terme, qu'il en est conscient et que le terme recherché est imminent, d'où l'expression, 'sur le bout de la langue' (Brown & McNeill, 1966). Que le locuteur connaisse le terme est démontrable. Soit il le produit spontanément peu de temps après (en général dans la journée), reconnaissant par ailleurs, et souvent avec soulagement, que c'est bien le terme recherché, soit il l'identifie dans une liste, tâche qu'il effectue avec une vitesse et certitude étonnante (taux d'erreurs extrêmement faible).

D'autres particularités du MBL sont le fait que le locuteur, sait énormément de choses concernant le mot-cible bien qu'il soit incapable de le produire : fragments de *sens* ou fonctions pratiques ('cela sert à s'orienter lors d'une navigation en mer'), *informations syntaxiques* (catégorie lexicale : nom/verbe ; genre grammatical : masculin/féminin), *informations morphologiques* (type et origine de l'affixe) ; *informations phonologiques* (contour intonatif, nombre de syllabes, première et dernière syllabe).

Les informations données par les personnes se trouvant dans cet état ont souvent été utilisées par des psychologues comme argument pour construire et justifier un modèle de production lexicale. La plupart des chercheurs s'accordent pour dire qu'il y a deux étapes se succédant avec, ou sans, chevauchement (Levelt et al. 1999; Ferrand, 1998 ; mais voir également Caramazza, 1997). L'une consiste à déterminer le *lemme* (pour un sens donné on choisit une forme lexicale, qui elle est abstraite), l'autre a pour fonction de déterminer la forme concrète (forme morphologique, graphémique ou phonologique), le *lexème*. Si l'enchaînement de ces deux étapes se fait généralement en continu, donc sans interruption, des problèmes peuvent survenir. Ainsi, il se peut que la première étape se déroule correctement mais pas la seconde, auquel cas on aboutit à l'état nommé MBL : l'information sémantique et grammaticale étant disponibles intégralement, mais pas l'information graphique ou phonologique. Bien entendu, on peut aussi imaginer que l'accès sémantique soit déficient, auquel cas il est logiquement impossible d'accéder à la forme phonologique, car, à moins de ne répéter un mot, il est impossible d'avoir accès à sa forme phonologique sans en avoir déterminé le sens.

Etant donné la faiblesse de la trace phonologique on pourrait être tenté à vouloir renforcer celle-ci, et c'est bien ce que certains psychologues ont suggéré (Abrams et al., 2007). Pourtant, ce n'est pas la voie que nous allons emprunter ici et il y a pour cela plusieurs raisons.

L'analyse d'erreurs (Rossi, 2001)[4] et l'étude du phénomène du MBL suggèrent que l'accès lexical se fait par deux voies : par le sens et par la forme (notamment les sons, phonèmes). Ceci dit, l'accès par le sens (boisson-vin) n'exclut nullement l'accès par des termes associés, par exemple, le terme 'vin' activant le terme 'fromage'. Pourtant, le terme 'vin' n'est pas un élément de sens du terme 'fromage'. C'est une co-occurrence, ou si l'on préfère, c'est une association sur l'axe syntagmatique. Outre cette co-occurrence, les deux termes entretiennent une relation sémantique (ou encyclopédique : 'en mangeant du fromage on boit du vin'). Deux autres points méritant être rappelés sont l'aspect relationnel des termes (ils sont du type associatifs : un terme *x* pouvant évoquer un terme *y* avec une probabilité *z*) et leur organisation sous forme de graphe. Ces deux caractéristiques sont capitales, et elles offrent plusieurs avantages :

- le fait que des termes soient liés élargit le champ de recherche : chaque terme source (mot donné en entrée) active un ensemble de termes associés (termes cibles potentiels), ensemble susceptible de contenir le terme recherché;
- le fait que les termes soient organisés sous forme de graphe permet leur accès par différents chemins. Si cette forme de représentation introduit une certaine redondance dans la représentation des données, elle a l'immense mérite de permettre de retrouver le bon chemin, au cas où l'on se serait trompé à un certain embranchement, situation guère possible, ou du moins beaucoup plus compliquée, dans le cas d'arborescences.

---

causer cette forme de blocage. Produire un mot suppose avoir effectué des traitements à différents niveaux (conceptuels, sémantiques, phonologiques). Or, l'échec (erreur, incomplétude) à n'importe lequel de ces niveaux peut bloquer la machine et produire ce qu'on appelle *manque de mot*. Le terme MBL ne décrit qu'une situation très particulière : le blocage se situe uniquement au niveau phonologique (informations erronées, informations manquantes), les informations venant des autres niveaux étant généralement disponibles dans leur intégralité.

[4] Des erreurs comme 'à ma gauche' au lieu de 'à ma droite' et 'élision' à la place de 'illusion' illustrent ces deux voies d'accès.

En somme, lorsqu'on est en état de MBL on peut essayer de retrouver un terme via d'autres termes phonologiquement proches (accès par la forme, Abrams, 2007; Zock; 2002), mais aussi via des termes ayant un lien sémantique. Nous nous sommes intéressés ici uniquement à cette dernière solution.

## 2.1    Hypothèses de travail

Il semble difficile de distinguer les deux usages suivants d'une application de MBL : 1) l'utilisation comme un utilitaire afin de retrouver un terme, 2) l'usage ludique de type devinette. Les motivations pour le second usage sont variables, mais en général visent à «mettre en difficulté le système ».

Il semble *a priori* difficile de savoir si les utilisateurs abordent notre outil de façon utilitaire ou ludique. Plutôt qu'effectuer une étude approfondie sur cette question (étude sans doute longue, difficile et coûteuse), nous allons donc considérer ces deux activités comme identiques. Plus précisément, nous partons des deux hypothèses suivantes :

- Hypothèse 1 : les termes recherchés par les utilisateurs de notre jeu MBL sont des termes de fréquence moyenne ou faible (termes de difficulté[5] moyenne ou importante). Les utilisateurs sont vraiment intéressés à trouver le terme recherché.
- Hypothèse 2 : le vocabulaire ciblé est de basse et de moyenne fréquence (termes de difficulté moyenne ou importante). Les joueurs cherchent à vérifier l'efficacité de l'outil.

Etant donné que le vocabulaire qui déclenche le MBL et celui avec lequel les joueurs de MBL jouent sont identiques, cela nous amène à postuler que « *l'évaluation d'un outil de MBL peut se faire grâce à des joueurs* ».

Par ailleurs, une seconde hypothèse de travail consiste à dire que l'éventail de comportements des joueurs est comparable à celui des personnes ayant réellement besoin de retrouver un mot. A savoir, leur motivation consiste à essayer de piéger l'outil soit avec un terme simple et des indices à la marge, soit avec un terme rare, improbable, ou récent, et des indices plus directs. On peut donc raisonnablement conclure qu'une telle évaluation est plus défavorable que celle portant sur des cas réels de MBL et qu'elle caractérise une ligne basse : *l'évaluation d'un outil de MBL via un jeu fournit une valeur plancher de ses performances*.

Ce sont ces deux hypothèses que nous tenterons de vérifier dans la suite de cet article.

## 3    Constitution du réseau lexical

### 3.1    JeuxDeMots[6] : construction du réseau

Le principe de base conduisant grâce à un jeu en ligne à la construction progressive du réseau lexical, à partir d'une base de termes préexistante, a déjà été décrit par (Lafourcade et Joubert, 2009). Une partie se déroule entre deux joueurs, en double aveugle et en asynchrone. Pour un même terme cible T et une même consigne C (synonymes, domaines, association libre…), les deux joueurs proposent des termes correspondant, selon eux, à cette consigne C appliquée à ce terme T. Ces propositions sont limitées en nombre, ce qui a pour effet d'augmenter leur pertinence, mais également dans le temps pour favoriser leur caractère spontané. Nous mémorisons alors les réponses communes à ces deux joueurs[7]. Les validations sont donc faites par concordance des propositions entre paires de joueurs pour un même couple (C,T). Ce processus de validation rappelle celui utilisé par (von Ahn et Dabbish, 2004) pour l'indexation d'images ou plus récemment par (Lieberman et al., 2007) pour la collecte de « connaissances de bon sens ». À notre connaissance, il n'avait jamais été mis en œuvre dans le domaine de la construction des réseaux lexicaux.

---

[5]    Nous faisons également l'hypothèse que ce que nous appelons la difficulté d'un terme est contra-variante à sa fréquence, la difficulté d'un terme exprimant à la fois la difficulté à trouver des indices s'y rapportant mais également la difficulté à faire émerger ce terme chez un interlocuteur.

[6]    http://jeuxdemots.org

[7]    La limitation dans le temps de la saisie des propositions des joueurs peut favoriser les fautes d'orthographe, mais, comme nous ne mémorisons que les réponses communes aux deux joueurs d'une même partie, l'expérience montre que ce risque est très limité et que seules subsistent les fautes d'orthographes qui de toutes façons auraient été faites par les joueurs (par exemple : *beau* pour *bot*, en parlant de *pied*).

La structure du réseau lexical ainsi obtenu s'appuie sur les notions de nœuds et de relations entre nœuds, selon un modèle initialement présenté par (Collins et Quillian, 1969) et davantage explicité par (Polguère, 2006). Chaque nœud du réseau est constitué d'une unité lexicale (terme, raffinement d'un terme ou segment textuel) liée aux autres termes via des relations des fonctions lexicales, telles que présentées par (Mel'čuk et al., 1995). Les relations obtenues grâce à l'activité des joueurs sont typées et pondérées[8] : elles sont typées par la consigne imposée aux joueurs, elles sont pondérées en fonction du nombre de paires de joueurs qui les ont proposées, comme indiqué dans (Lafourcade et Joubert, 2009). Plusieurs exemples de relations acquises ont été donnés dans (Lafourcade et Joubert, 2009). Au moment du lancement de JeuxDeMots en juillet 2007, le réseau comportait 152 000 termes (non reliés entre eux, c'est-à-dire aucune relation n'existait). Courant mars 2011, à l'issue d'environ 900 000 parties jouées par plus de 2500 joueurs, notre réseau compte 229 000 termes et plus de 1 100 000 relations.

### 3.2   PtiClic : consolidation du réseau

De manière analogue à JeuxDeMots (JDM), une partie de PtiClic (http://pticlic.org) se déroule, en double aveugle et asynchrone, entre deux joueurs. Un premier joueur se voit proposer un terme cible T, origine de relations, ainsi qu'un nuage de mots provenant de l'ensemble des termes reliés à T dans le réseau lexical produit par JDM. Plusieurs consignes correspondant à des types de relations sont également affichées. Le joueur associe, par cliquer-glisser, des mots du nuage aux consignes auxquelles il pense qu'ils correspondent. Ce même terme T, ainsi que le même nuage de mots et les mêmes consignes, sont également proposés à un deuxième joueur. Selon un principe analogue à celui mis en place pour JDM, seules les propositions communes aux deux joueurs sont prises en compte, renforçant ainsi les relations du réseau lexical.

Contrairement à JDM, PtiClic est un jeu fermé où les utilisateurs ne peuvent pas proposer de nouveaux termes, mais sont contraints de choisir parmi ceux affichés. Ce choix de conception a pour but de réduire le bruit dû aux termes mal orthographiés ou aux confusions de sens. PtiClic réalise donc une consolidation des relations produites par JDM et permet de densifier le réseau lexical. Notons également que PtiClic permet de créer de nouvelles relations entre termes précédemment reliés par au moins une relation d'un autre type (même si ce n'est pas l'objectif premier de ce logiciel). Afin de réduire le silence correspondant aux termes non proposés par les utilisateurs de JDM, (Zampa et Lafourcade, 2009) ont suggéré de générer le nuage de mots à l'aide de la LSA, en utilisant un corpus externe de grand volume (l'expérimentation réalisée utilise un corpus comportant une année du journal « Le Monde ».). Cette solution permet d'augmenter le réseau lexical par ajout de nouvelles relations, en proposant aux joueurs de nouveaux termes cibles sans liens à T dans le réseau.

### 3.3   Raffinement des termes : enrichissement du réseau

Le processus permettant d'aboutir aux raffinements de termes est décrit dans (Lafourcade et Joubert, 2010). Nous avons fait l'hypothèse que les sens d'usage, plus communément appelés usages, d'un terme correspondent dans le réseau aux différentes cliques auxquelles ce terme appartient. Cette approche est analogue à celle développée par (Ploux et Victorri, 1998) à partir de dictionnaires de synonymes. En calculant la similarité entre les différentes cliques d'un même terme, nous pouvons construire son arbre des usages nommés. La racine de l'arbre regroupe tous les sens de ce terme. Plus on s'éloigne de la racine, c'est-à-dire plus la profondeur des nœuds dans l'arbre est importante, plus on rencontre des distinctions fines d'usages. Les nœuds de profondeur 1 dans cet arbre correspondent généralement aux différents sens de ce terme répertoriés dans les dictionnaires traditionnels. Après un processus de validation par un expert lexicographe de ces différents sens, nous les intégrons dans le réseau en tant que nœuds de raffinement du terme considéré ; le réseau est ainsi enrichi de nouveaux nœuds à partir desquels ou vers lesquels les joueurs de JDM peuvent créer des relations. Actuellement, sur les 229 000 termes connus par le réseau, près de 5 000 ont été raffinés.

---

[8]   Une relation peut donc être considérée comme un quadruplet : terme source, terme cible, type et poids de la relation. Entre deux mêmes termes, plusieurs relations de types (et de poids) différents peuvent exister.

## 4   Un algorithme et un outil de MBL

AKI est un outil d'accès lexical accessible sur le Web à partir du portail JeuxDeMots ou directement à http://www.lirmm.fr/jeuxdemots/AKI.php. AKI peut être envisagé comme un jeu : l'utilisateur fait ″deviner″ un terme cible à l'ordinateur (espérant, éventuellement, de manière secrète, de mettre en défaut sa capacité à trouver un terme à partir d'indices). AKI peut également être considéré comme une assistance, pour retrouver un terme qu'on a sur le bout de la langue. L'utilisateur est invité à fournir, un par un,  une succession de termes indices qui lui paraissent pertinents pour trouver le terme cible recherché. Ce mécanisme est comparable à celui de certains jeux télévisés. Après chacun de ces termes indices AKI fait une proposition. Si elle correspond au terme recherché, l'utilisateur valide la proposition, sinon il introduit un nouvel indice. Ce dialogue se poursuit jusqu'à ce que l'une des deux situations se produise : AKI trouve le terme cible ou il abandonne et demande à l'utilisateur de fournir la solution.

### 4.1   AKI : principe et réalisation

L'utilisateur saisit un premier terme indice $i_1$. En utilisant le réseau lexical, l'algorithme calcule la signature lexicale de $i_1$ : $S(i_1) = S_1 = t_1, t_2, \ldots$ où les $t_i$ sont triés par activation décroissante. Autrement dit, $t_1$ est le terme pour lequel la somme des poids des relations le liant à $i_1$ est la plus élevée. La première proposition d'AKI, $p_1$, correspond à ce terme : $p_1 = t_1$. Si c'est le terme cible, l'utilisateur le valide et la partie est terminée, sinon, il est retiré de la signature $S_1$, ainsi que $i_1$ (qui ne peut pas être le terme cible). Donc, à ce stade, la signature courante est : $S'_1 = S_1 - \{p_1, i_1\}$. L'utilisateur est alors invité à saisir un deuxième terme indice $i_2$. L'algorithme calcule une deuxième signature lexicale par intersection entre la signature courante et celle de $i_2$ : $S_2 = (S'_1 \cap S(i_2)) - i_2$.

Le terme proposé $p_2$ est celui de $S_2$ dont l'activation est la plus forte. Autrement dit, parmi les termes reliés à la fois à $i_1$ et à $i_2$, AKI affiche celui pour lequel la somme des poids des relations le reliant à $i_1$ et à $i_2$ est la plus élevée, exception faite des termes déjà proposés par AKI pour cette partie (ainsi que des termes indices !). La signature courante est alors $S'_2 = S_2 - p_2$. D'une façon générale, à l'étape *n*, nous avons :

$$S_n = (S'_{n-1} \cap S(i_n)) - i_n \quad \text{et} \quad S'_n = S_n - p_n$$

où $i_n$ est le n-ième indice fourni par l'utilisateur et $p_n$ la n-ième proposition de AKI. Le nombre de termes constituant la signature diminue donc au fur et à mesure de l'insertion d'indices. Il est fréquent que la signature devienne vide, avant même que le terme cible n'ait été trouvé ; dans ce cas là, AKI ne peut plus proposer de terme. Le processus pourrait s'arrêter là, mais afin d'améliorer le taux de rappel, une ″procédure de rattrapage″ est mise en œuvre : au lieu d'effectuer des intersections de signatures, on utilise leur somme :

$$S_n = (S'_{n-1} + S(i_n)) - i_n \quad \text{et} \quad S'_n = S_n - p_n$$

Cette procédure favorise l'apprentissage en créant des relations entre des termes isolés. Aussi utile soit elle, cette astuce doit néanmoins être utilisée avec précaution. En effet, au-delà de quelques itérations, le nombre de termes constituant la nouvelle signature devient vite prohibitif ; notre expérience tend à montrer qu'il ne faut pas dépasser deux itérations. Au delà, l'algorithme donne des propositions trop éloignées des mots proposés. Le processus se termine lorsque AKI a trouvé le terme cible recherché, ou lorsque la signature lexicale courante devient vide (ce qui est relativement rare, compte tenu de la procédure de rattrapage). À partir de 5 indices (cette limite de 5 étant un paramètre modifiable) l'utilisateur a la possibilité d'abandonner en indiquant à AKI qu'il fait fausse route. En effet, nous avons estimé que si, au bout de 5 indices, AKI n'a pas trouvé le terme recherché, cela signifie probablement que ces indices ne sont pas pertinents. La figure 1 présente quelques exemples de parties.

Les utilisateurs d'AKI ont la possibilité de faire précéder leur indice de mot-clé faisant référence à des fonctions lexico-sémantiques. Actuellement il est possible d'utiliser dix fonctions : hyperonymie, hyponymie, synonymie, antonymie, domaine, matière, lieu (lieu typique où l'on peut trouver ce que l'on cherche), caractéristique, holonymie, méronymie. Elles correspondent toutes à un type de relation existant dans le réseau JeuxDeMots.

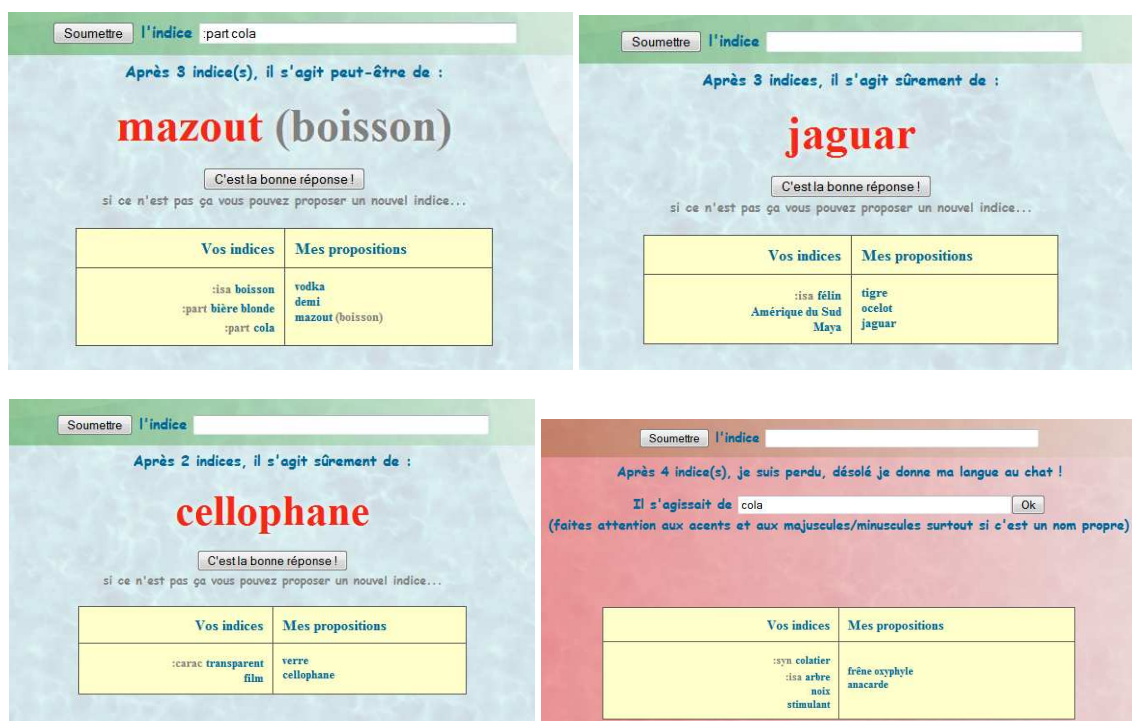

Figure 1 : Quelques exemples de parties. Dans les trois premiers cas, AKI a trouvé le terme cible. Dans le quatrième cas, ne pouvant plus faire de proposition, AKI a abandonné et l'utilisateur a saisi la bonne réponse (il s'agissait donc d'une utilisation ludique de AKI).

### 4.2 Consolidation du réseau

Dans l'hypothèse jeu, quand le terme cible n'a pas été trouvé par AKI, l'utilisateur est invité à le saisir. Il y a alors création dans le réseau lexical de relations typées « AKI » avec un poids très faible (+1 à chaque occurrence). Ces relations sont régulièrement vérifiées et validées (ou non) par un expert lexicographe. En effet, dans la mesure où c'est l'utilisateur lui-même qui choisit le terme cible, ainsi que les termes indices, la sécurité de la pertinence de telles relations peut difficilement être garantie (un joueur peut toujours commettre des erreurs, sans parler d'éventuels joueurs malveillants). Ceci est différent de JeuxDeMots où les relations sont créées par intersections de propositions de joueurs. Ici, il arrive fréquemment qu'un utilisateur joue plusieurs fois le même terme, avec des indices différents mais également avec des indices communs. Nous sommes en train de réfléchir comment sécuriser ces relations typées « AKI ».

## 5 Evaluation du réseau via AKI

**Évaluation informelle -** Nous avons mené une évaluation informelle des performances d'AKI à partir du jeu de société *Tabou* inversé. Le principe du jeu de société *Tabou* est de faire deviner un terme à des personnes à l'aide d'indices, en excluant certains termes dits tabous. La version commerciale de ce jeu fournit une collection de 500 fiches comprenant chacune un terme cible et 5 termes tabous. La version inversée de ce jeu consiste à faire deviner le mot cible en énumérant ces termes tabous, l'hypothèse étant que ces termes sont particulièrement évocateurs du terme cible lorsqu'ils apparaissent ensemble.

Ceci étant, nous avons soumis cette collection de 500 fiches à AKI ainsi qu'à trois personnes (à des fins de comparaisons). AKI a retrouvé le terme cible au plus tard au bout des 5 indices dans 494 cas (soit 98,8% de réussite). Les personnes prises comme références, ont globalement trouvé (dans les mêmes conditions) 402 fois (soit 80,4% de réussite). Ce dernier chiffre n'est qu'une indication, vue la faible taille de l'échantillon considéré, de trois participants.

### 5.1 Protocole

L'évaluation, tout comme l'apprentissage, ne se fait qu'en fonction de ce que les joueurs ont renseigné. Elle se fait donc sur du vocabulaire appartenant à la classe ouverte. Comme déjà mentionné, AKI peut être envisagé comme un jeu ou un outil de MBL. *A priori*, notre logiciel ne sait pas faire la distinction entre les deux usages. En effet, sur une seule partie et si AKI trouve la solution, nous ne pouvons pas savoir a priori si l'utilisateur connaissait ou s'il recherchait le terme cible. Par contre, si dans un laps de temps relativement court (de l'ordre de quelques minutes) un même terme est joué plusieurs fois, nous pouvons faire l'hypothèse qu'il s'agit d'une utilisation de type jeu (au moins à partir de la deuxième partie) où l'utilisateur essaie de faire trouver le terme cible par AKI, en proposant généralement des indices différents. Dans chacun des deux cas, jeu ou outil de MBL, les termes cibles sont majoritairement des termes de fréquence moyenne, voire faible. En effet, jouer pour trouver un mot fréquent ne présente pas un grand intérêt, et généralement on ne recherche pas grâce à un outil de MBL un terme courant. Les graphiques ci-après montrent l'évolution dans le temps du rapport entre le nombre de parties gagnées par AKI, parties où l'utilisateur a indiqué que le logiciel a trouvé le terme cible, et le nombre de parties jouées. Les graphiques de cette section reflètent 6522 parties réalisées entre le 30/12/2010 et le 24/01/2011.

### 5.2 Analyse quantitative et évolution dans le temps

Le premier graphique (figure 2) présente l'évolution du rapport entre le nombre de parties gagnées et le nombre de parties jouées par fenêtre glissante de 500. Par exemple, à l'abscisse 100 (correspondant à 2000 parties), la courbe correspond à la moyenne des valeurs entre les parties 1501 et 2000. Lorsqu'il y a moins de 500 valeurs, la courbe présente la moyenne des *n* premières valeurs. Ce graphique montre globalement une légère amélioration des résultats au cours du temps, avec un passage de 60% de réussite à 80%.

Nous avons analysé le type de mots joués par les utilisateurs. Nous avons considéré comme **courants** les mots issus de **l'échelle orthographique Dubois Buyse**[9], c'est-à-dire, ceux connus par un enfant de 12 ans. Nous considérions les autres comme normaux. Par exemple, *fourchette, pie, écureuil, restaurant* sont courants, tandis que *séquoia géant, Rabat* ou même *Akinator* sont considérés comme des termes normaux. On peut considérer que globalement, les termes normaux ont une fréquence d'utilisation allant de moyen à rare (les mots courants ayant une fréquence d'utilisation élevée). L'analyse des parties jouées, ainsi que le nombre de mots différents nous révèle que dans les deux cas, environ 25% concerne des mots courants : sur 1701 mots différents joués, 435 sont courants (soit 25,6%) et sur 6488 parties, 1565 concernent des mots courants (24,1%). Il est important de noter que les mots sont ceux qui étaient déjà bien complets dans le réseau lexical de JDM. On remarquera que l'addition pondérée par le nombre de parties des deux courbes (figures 3 et 4) donne la courbe de la figure 2.

Sur les parties jouées sur des mots courants, on observe globalement une stagnation des résultats, preuve que le réseau était déjà bien complet. Ce qui n'exclut pas que le réseau ait été enrichi de nouveaux résultats, bien que ceux-ci soient très peu visibles. En revanche, en ce qui concerne les mots normaux (ceux qui ne sont pas considérés comme courants), la progression est bien plus claire. Alors qu'au départ, la moyenne à 1000 était inférieure à 60%, elle atteint 80% à la fin. À quoi est due cette progression ? Une première explication possible serait que les joueurs découvrent AKI ; et ce n'est que petit à petit qu'ils réussissent à proposer des indices pertinents. Ceci est quelque peu contredit par l'expérience : en effet, il semble que les joueurs essaient de plus en plus de proposer des indices indirects afin de « mettre le système en défaut ». Il nous parait plus plausible que cette progression serait due à la capacité d'apprentissage du système. Cette hypothèse serait bien entendu à vérifier sur un plus long terme mais le nombre de relations acquises lors de cette expérience semble la corroborer.

---

[9] L'échelle orthographique Dubois Buyse permet d'indiquer les mots normalement acquis par 75% des enfants d'une classe d'âge. Nous considérons donc comme vocabulaire courant les mots bien orthographiés par 75% des enfants de 12 ans. On peut trouver cette liste, entres autres, à http://o.bacquet.free.fr/db2.htm

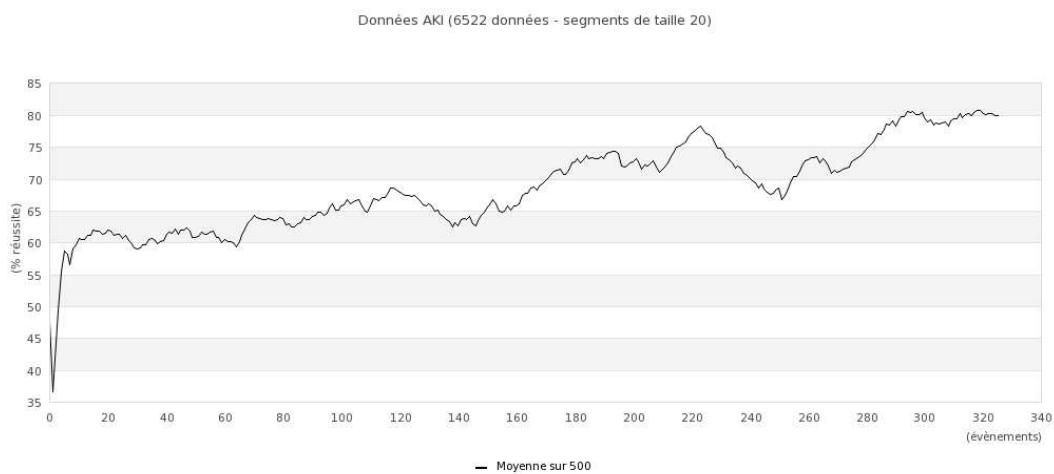

Figure 2 : Graphique montrant l'évolution dans le temps du rapport entre le nombre de parties AKI gagnées et le nombre de parties jouées (moyenne glissante sur les 500 dernières parties jouées)

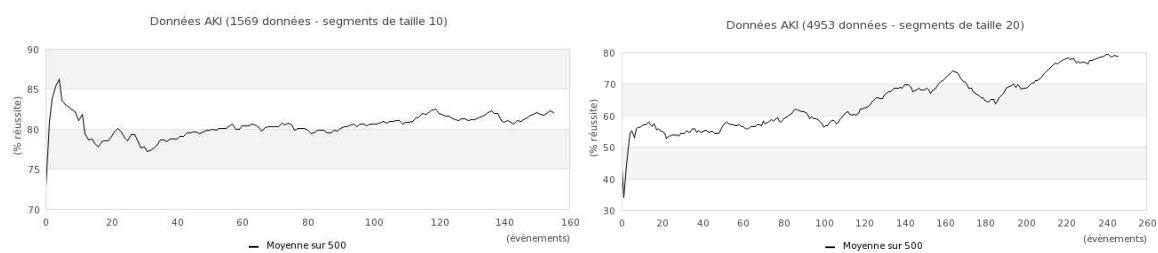

Figure 3 et 4 : A gauche, l'évolution dans le temps du rapport entre le nombre de parties gagnées et le nombre de parties jouées sur des **mots courants** (moyenne glissante sur les 500 dernières parties jouées) - A droite, l'évolution dans le temps du rapport entre le nombre de parties gagnées et le nombre de parties jouées sur des **mots de fréquence moyenne à faible** (moyenne glissante sur les 500 dernières parties jouées)

**Acquisition de termes.** Depuis le 1er janvier 2011, 208 nouveaux termes ont été insérés dans le réseau lexical via AKI. Ils résultent de 724 parties jouées. La quasi-totalité (90%) de ces termes sont des entités nommées (*DCRI, Révolution de jasmin, Noob, …*), ou (10%) des termes composés et des néologismes divers (*sexe par surprise, gaz lacrymogène,…*), souvent liés à l'actualité.

**Acquisition de relations.** Depuis le 1er janvier 2011, 11434 relations on été acquises à l'aide de AKI (6546 parties). Si on ne compte que celles absentes du réseau, ce nombre tombe à 2105. Donc, en moyenne le réseau acquiert 1 nouvelle relation toutes les trois parties.

### 5.3 Analyse qualitative des parties

**Sur le type de vocabulaire.** Le vocabulaire (après une première analyse) se découpe en nombre de parties jouées en 24% de mots courants, le reste se divisant en 50% de mots de fréquence moyenne ou faible, et 26% de termes souvent nouveaux et liés à l'actualité. On peut considérer que ce dernier groupe est à rapprocher des 50% si on partitionne les termes entre mots courants et les autres. Les termes liés à l'actualité conduisent souvent (69%) à un échec ce qui semble normal, étant donné qu'il s'agit de termes nouveaux (par exemple : *Révolution de jasmin, Jean-Luc Mélenchon, Médiator* - essentiellement des entités nommées) ou de termes déjà connus, mais recherchés via des indices nouveaux (*président, Tunisie, fuite => Ben Ali*). Certains mots déjà connus de AKI, c'est-à-dire présents dans le réseau lexical, sont réactualisés par l'actualité : *trafic d'organes, Lance Armstrong, aspartame*. Le compte sur l'ensemble des termes joués (indépendamment du nombre de parties) donne environ la même répartition de 1/4 de vocabulaire courant et de 3/4 de termes rares ou récents.

**Sur les indices proposés.** Le nombre moyen d'indices pour trouver un mot est de 2,8. Dans 40% des cas, un terme courant est trouvé dès le premier indice. Un peu moins de 3% des parties sont poursuivies au-delà de 5 indices, les 97% se divisant entre les trois cas suivants : a) le mot est trouvé avant, b) AKI échoue avant, ou c) l'utilisateur abandonne. Quand la recherche va au delà de 5 indices, la partie aboutit à une réussite dans 60% des cas. Il s'agit de termes de domaine fortement lexicalisés, et AKI est sur la bonne voie.

Une analyse des indices donnés lors des parties indique que moins de 0,3% des parties comporte au moins un indice apparemment non cohérent. Soit l'utilisateur a voulu volontairement mettre le système en défaut en donnant un indice sans rapport, soit il s'agit d'une erreur ou d'une confusion. On remarquera (avec satisfaction) que la quasi totalité des parties sont jouées "honnêtement" (ce qui peut s'expliquer par le manque d'intérêt à mettre en défaut le système avec des indices absurdes). On peut grouper les indices proposés en deux catégories :
- les indices **frontaux** (noté F) sont ceux qui amènent rapidement à la solution (trois indices au maximum). Dans le réseau, ils sont fortement connectés à la solution, en général de façon bidirectionnelle. Par exemple : *félin* pour *chat*.
- les indices **latéraux** (notés L) sont ceux qui sont très faiblement connectés à la solution et par ailleurs beaucoup plus fortement connectés à d'autres termes. Par exemple : *lisse, blanc, froid* pour *lavabo*.

Les parties concernant les mots courants correspondent à des parties dont la séquence type est : L+ (une succession d'indices latéraux). Plus le terme cible devient rare ou récent, plus la séquence type se rapproche de F+ (une succession d'indices frontaux). Il existe quelques autres schémas de parties, mais qui restent fort minoritaires. Les schémas les plus notables sont :
- L+ : uniquement des indices latéraux : *garçon, caillou, oiseau, miette* pour *Le Petit Poucet*.
- F+ : uniquement des indices frontaux : *sang* ou *couleur* pour *rouge* ou encore *mammifère, marin, défense* pour *morse*.
- L+, F : une série d'indices latéraux puis un dernier indice de type frontal permettant de trouver la solution *: blanc, lisse, dur, éléphant* pour *ivoire*.

On peut raisonnablement supposer que le schéma L+, F correspond à une activité ludique, plutôt qu'à une activité utilitaire. C'est sans doute également le cas pour le premier schéma, lorsqu'il s'agit de termes fréquents.

**Sur le type d'activité.** Pouvons nous déduire de l'activité enregistrée qu'il s'agit d'une activité ludique ou d'un usage MBL ? Sans doute seulement partiellement. En revanche, de nombreux indices nous portent à croire que la plupart des parties enregistrées lors de notre expérience relèvent du jeu. Nous savons que les liens partagés par nous ou des joueurs des réseaux sociaux Facebook et Twitter[10], ont généré 60% de l'activité de AKI. Ces liens proposaient de jouer tel ou tel mot. L'envoi du lien vers AKI à des listes de diffusion professionnelles (laboratoires, enseignements, sociétés savantes) générait une forte augmentation du trafic (pratiquement les autres 40%). À moins de ne penser que toutes ces personnes cherchaient le même mot à ce moment précis, on peut supposer que l'immense majorité de l'activité de AKI relève du jeu. Toujours pour aller dans ce sens, nous n'avons eu directement que deux témoignages attestant une utilisation non ludique ; dans ces deux cas, AKI s'est révélé fort utile puisqu'il a donné satisfaction aux utilisateurs.

---

[10] Pour voir en temps réel le compte des parties d'AKI, http://twitter.com/#!/Tot_aki

### 5.4 Conclusion de l'évaluation

À l'aune des résultats ci-dessus, il ne nous est pas permis de conclure avec certitude que les hypothèses 1 et 2 présentées au début de cet article sont globalement valides, mais elles ne sont pas invalidées pour autant. De nombreux indices nous laissent penser que quasiment toutes les parties analysées proviennent du jeu et non d'une utilisation réelle. Nous avons également montré que le vocabulaire utilisé durant ces jeux était le même vocabulaire que celui faisant l'objet du MBL. Ceci permet de valider notre première hypothèse de travail, à savoir, que *l'évaluation d'un outil de MBL peut se faire grâce à des joueurs.* En revanche, notre seconde hypothèse de travail —(*l'éventail des comportements des utilisateurs jouant avec un outil de MBL inclut le comportement de ceux ayant réellement besoin de retrouver un mot*)— demanderait une analyse plus fine.

Le réseau et sa consolidation via l'activité générée avec AKI permettent, dans le cas de vocabulaire complètement ouvert, de trouver le terme dans 78% des cas. Dans le cas de vocabulaire considéré comme courant, on se situe aux alentours de 82%. Enfin, dans le cas de vocabulaire filtré (issu du jeu Tabou inversé), on atteint 98,8%. On notera que, dans ce dernier cas, la performance des êtres humains se situe aux alentours de 80%.

Nous avons évalué les performances de cinq personnes sur du vocabulaire tout venant. A cette fin, nous avons choisi au hasard pour chacun d'eux 100 termes parmi ceux joués dans AKI et pour lesquels ce dernier avait soit majoritairement trouvé (50 termes) soit échoué (50 termes). Les indices donnés étaient les 5 termes les plus fortement associés dans le réseau. La performance globale a été de 46%, chiffre à comparer avec les 75-80% d'AKI.

## Conclusion

Nous avons construit un réseau lexical évolutif de grande taille grâce à l'activité d'utilisateurs jouant en ligne (projet JeuxDeMots). Ces joueurs n'étant *a priori* pas des spécialistes, ce réseau représente un ensemble de connaissances générales communes. Avec des joueurs experts, il serait envisageable d'étendre ces connaissances, et donc le réseau, à des domaines spécialisés (n'est-ce pas déjà en partie le cas ?). L'intervention d'experts lexicographes, limitée à certaines corrections ainsi qu'à la validation des raffinements de termes, est ″négligeable″ compte tenu de la taille du réseau. Les questions concernant l'évaluation de la qualité d'une telle ressource, celles concernant son utilité, et la forme que peut prendre cette évaluation restent cependant ouvertes.

Le but poursuivi ici était d'évaluer la ressource lexicale ainsi produite à l'aide d'un logiciel (dénommé AKI). Celui-ci peut être considéré soit comme un jeu, soit comme un outil de MBL avec une approche exclusivement sémantique et lexicale. A l'heure actuelle, il n'y a aucune prise en compte de facteurs morphologiques ou phonologiques. AKI permet donc une évaluation à grande échelle du réseau par les utilisateurs eux-mêmes, qu'ils soient ou non des joueurs ayant contribué via JeuxDeMots. Quel que soit l'ensemble des termes considérés (termes courants ou termes de fréquence plus réduite) les performances d'AKI sont d'environ 80% ± 5%. Les résultats montrent par ailleurs que AKI est réellement utile, permettant de trouver des termes dans une ressource existante, tout en étant susceptible de l'enrichir grâce à sa capacité d'apprentissage.

On peut déduire des performance de AKI que 75% des termes pour lesquels il a été sollicité sont bien indexés, en tout cas suffisamment bien pour permettre le bon choix en cas de désambiguïsation lexicale (avocat: profession vs. fruit). L'évaluation se poursuit au long cours et les participants cherchant constamment à mettre en défaut AKI renforcent l'indexation, mais également l'évaluation avec une sévérité croissante – les deux se compensant. Il y a un auto-ajustement des joueurs en faveur d'indices faisant partie de la longue traîne. Une question restant cependant ouverte est de savoir à quel taux de réussite AKI va asymptotiquement plafonner. Cette valeur pourrait être un indice concernant une performance maximale en désambiguïsation lexicale en utilisant notre ressource.

## Références